\documentclass[11pt,twoside]{book} 
\usepackage{asp2008s}
\usepackage{times}
\usepackage{lscape}
\usepackage{epsf}

\usepackage{graphicx}
\usepackage{amssymb}
\usepackage{longtable}
\usepackage{rotating}
\usepackage{multirow}

\newcommand{\simless}{\mathbin{\lower 3pt\hbox {$\rlap{\raise 5pt\hbox{$\char'074$}}\mathchar"7218$}}}

\mainmatter

\newlength{\deftabcolsep}
\begin{document}

\title{The Canis Major Star Forming Region}


\author{Jane Gregorio-Hetem}


\affil{Universidade de S\~ao Paulo\\ Rua do Mat\~ao, 1226 - IAG/USP,
 05508-900, S\~ao Paulo, SP, Brazil}

\begin{abstract} 
The shape of the main arc formed by the Canis Major clouds has
been suggested to result from a supernova explosion possibly
triggering the recent star formation activity. The presence of dozens
of OB stars and reflection nebulae forms the CMa OB1/R1
associations. More than a hundred emission line stars are found in
this region, including the famous Z CMa, a binary system containing a
Herbig Be star and a FUor companion. Several embedded infrared
clusters with different ages are associated with the CMa clouds.
The main characteristics of the region in
terms of cloud structure, stellar content, age of associated young
clusters, distance, and X-ray emission are presented in this chapter.
Some of the arguments in favor
and against the hypothesis of induced star formation are discussed in
the last section.
\end{abstract}



\section{The Canis Major Associations}

The star-forming region in Canis Major is characterized by a
concentrated group of early type stars, which was first identified as
a stellar association by Ambartsumian (1947). The angular size of
$4^o$ in diameter was defined by Markarian (1952), who estimated a
distance of 960 ~pc, based on the identification of 11 probable
members.

Ruprecht (1966) established the approximate boundaries: $222^o < l <
226^o$ and $-3.4^o < b < +0.7^o$ of the CMa OB1 association. He
suggested that NGC~2353, a young cluster of age $\sim$ 12.6 ~Myr (Hoag
et al. 1961) and with a distance of $\sim$1315 ~pc (Becker 1963), constitutes the
main stellar concentration of CMa OB1.  More recently, Fitzgerald et
al. (1990) estimated an age of 76 ~Myr for NGC~2353 and a distance of 1.2~kpc, indicating that the cluster
is unrelated to CMa~OB1 despite their similar distances.

Clari\'a (1974a,b) studied the space distribution of O and B stars,
based on UBV photometry obtained for 247 stars.  The estimated E(B-V)
color excess confirms the existence of a group of young OB stars
together with excited gas and obscuring matter, which belong to the
CMa OB1 association.  Clari\'a used these data to derive a distance
of 1.15~kpc and an age of 3 ~Myr.

\begin{figure}[htp]
\centering
\includegraphics[draft=False,width=0.98\textwidth]{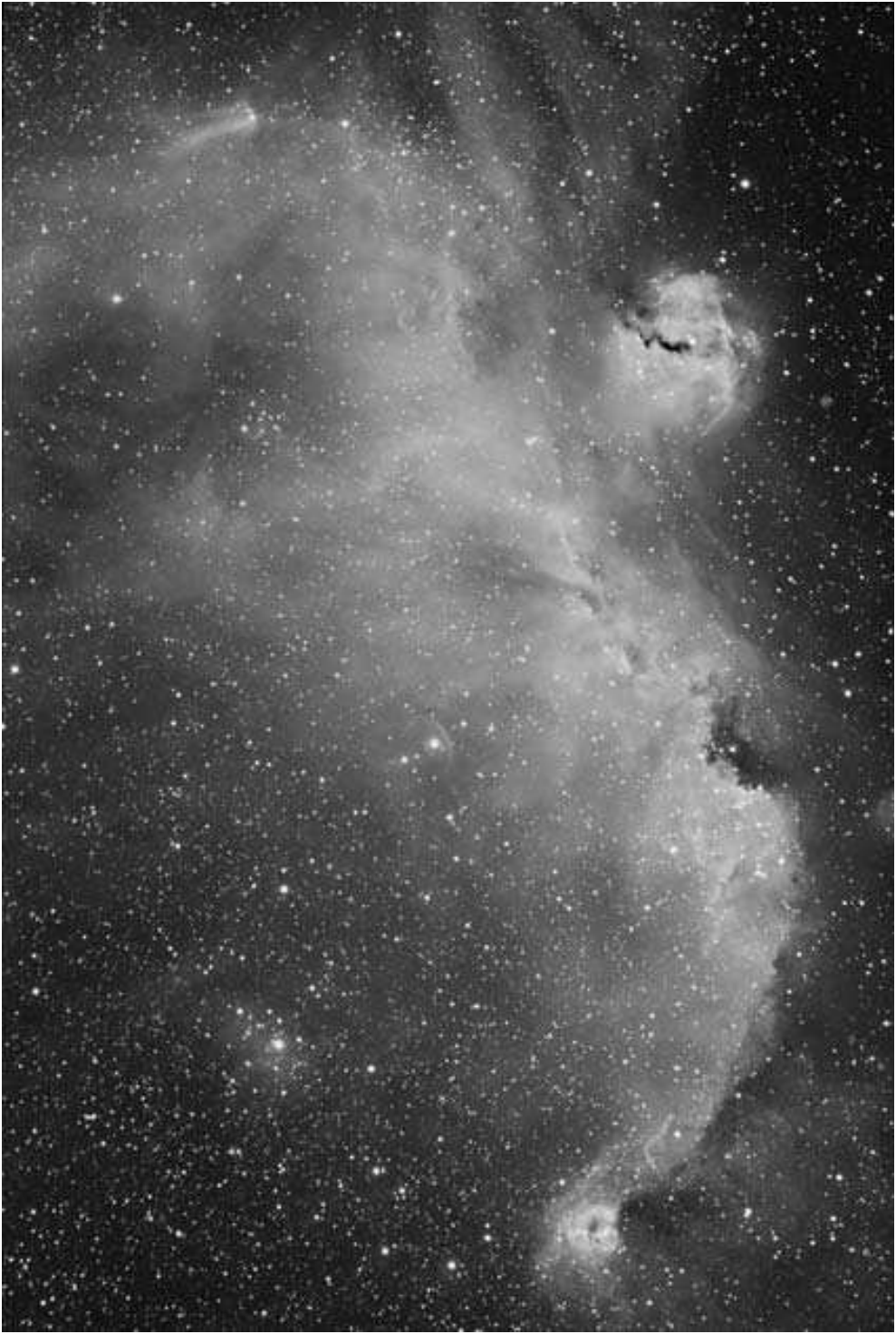}
\caption{The interface between the HII region and the neutral gas forms a
north-south oriented ridge in the CMa R1 association. The round
compact emission structure with a dark lane to the upper right is
IC~2177. Courtesy John P. Gleason.}
\end{figure}

\begin{figure}[!htb]
\vspace{-3mm}
\centering
\includegraphics[draft=False,width=0.9\textwidth]{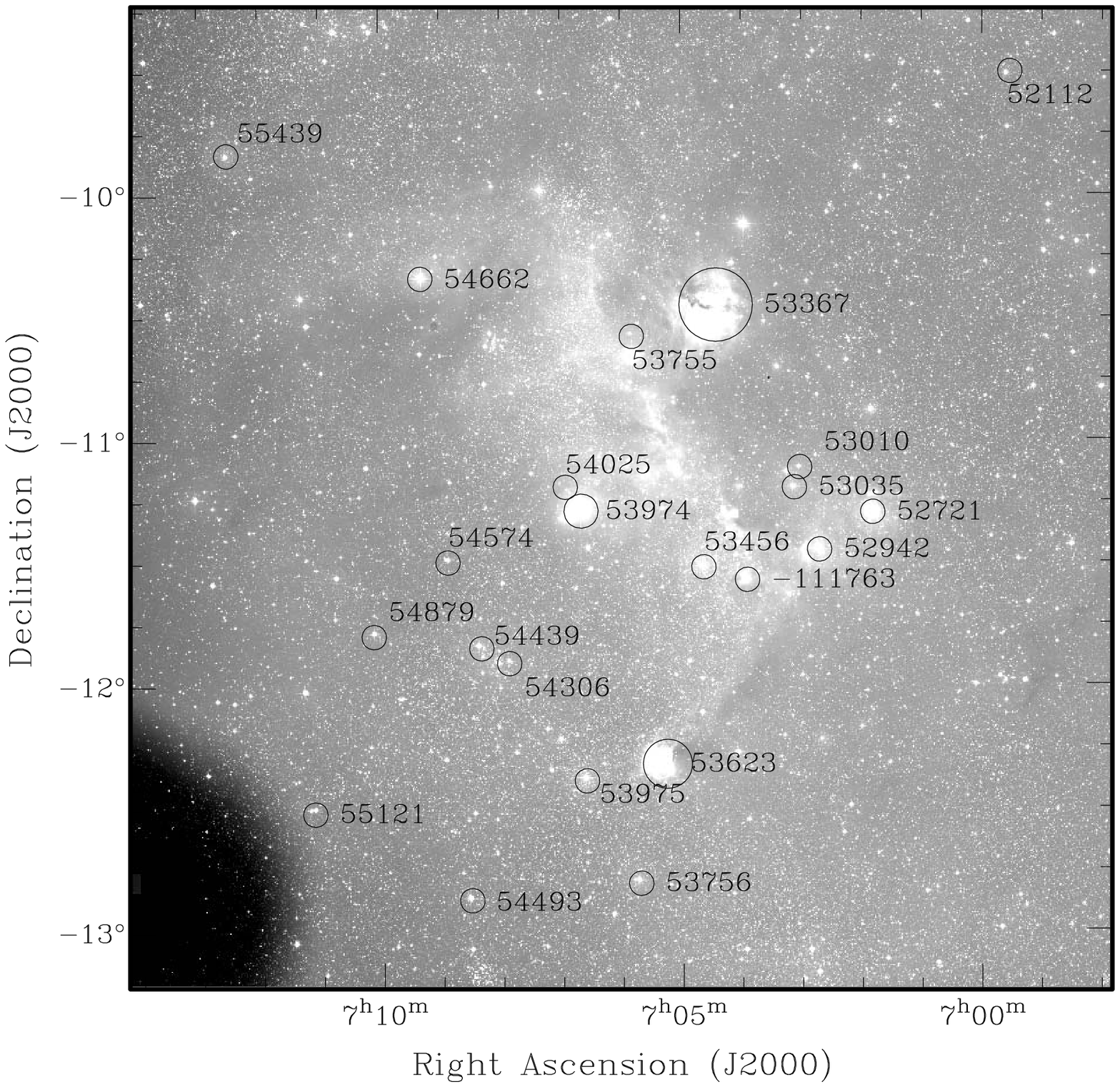}
\vspace{-3cm}
\caption{Optical image of CMa R1 obtained from the Digitized Sky Survey - DSS 2
in the R band (the dark-spot in the lower-right corner is an artifact). Some of the members are identified by their HD/BD numbers.}
\vspace{-3mm}
\end{figure}

The CMa R1 association is defined by a group of stars embedded in
reflection nebulae located within the boundaries of the OB1
association. It was first identified by van den Bergh (1966), who
listed ten stars distributed in the $-3.4^o < b < -2^o$ range. Racine
(1968) reported photometric and spectroscopic observations of the CMa
R1 members and suggested that the reflection nebulae seem to be nearer
than the CMa OB1 association. Clari\'a (1974b) compares the distance
moduli that he obtained with those values estimated by Racine. The evaluation
of the discrepancies led to the conclusion of the physical relation of the CMa OB1/R1 associations.

More than 30 nebulae (Herbst, Racine, \& Warner 1978) are found in the
CMa OB1 association, in particular three connected HII regions S292,
S296 and S297 (Sharpless 1959). Two of the conspicuous features are
the long arc of emission nebulosity S296 and the nearly circular
nebulosity IC~2177 surrounding the Be star HD~53367. The members of the
CMa OB1/R1 association are listed in Table 1. Figures ~1 and ~2 present
optical images of the entire region.
Schematic maps showing early-type stars, clusters,
emission and reflection nebulae, and dark clouds are presented in Figures
~3 and ~4.  Other associations or stellar clusters are found between $l\sim 233^o$
to $237^o$ and $b\sim -6^o$ to $-11^o$ (not identifiable in the figures
of this chapter). Kopylov (1958) denoted this
region as CMa II, or CMa OB2 in current nomenclature. Eggen (1981)
reported observations of 135 stars, which are members of the clusters
NGC ~2287 and Cr~121 (Collinder 1931). The distances derived from the
photometric data are 704 ~pc and 1.17 ~kpc, respectively. The age
determination indicates that NGC~2287 (100~ Myr) is considerably older
than Cr~121 (1.5~ Myr).



\begin{center}

{\small

\begin{longtable}{cccccccc@{\hskip5pt}c}

\caption{Members of CMa OB1/R1 presented by C74 (Clari\'a 1974b) and/or SEI99 (Shevchenko et al. 1999). The list gives Galactic ({\em l, b}) and equatorial (J2000) coordinates; visual magnitude and spectral type. Other cross identification names and related nebulae are also provided, when available - HRW (Herbst, Racine, \& Warner 1978); vdB (van den Bergh 1966) in the last column and Notes.} \\
\noalign{\smallskip}
\tableline
\noalign{\smallskip}
C74	&	SEI	&	HD/BD	&	{\em l, b}	&	RA	&	DEC	&	 V	&	ST & \\
        &       99        &               &               &    (J2000)    &      (J2000)  &      (mag)    &           & \\
\noalign{\smallskip}
\tableline
\noalign{\smallskip}
\endfirsthead

\caption{Members of CMa OB1/R1  (continued)} \\
\noalign{\smallskip}
\tableline
\noalign{\smallskip}
C74	&	SEI	&	HD/BD	&	{\em l, b}	&	RA	&	DEC	&	 V	&	ST & \\
        &       99      &               &               &    (J2000)    &      (J2000)  &      (mag)    &           & \\
\noalign{\smallskip}
\tableline
\noalign{\smallskip}
\endhead

\noalign{\smallskip}
\multicolumn{9}{l}{{Continued on Next Page\ldots}} \\
\tableline
\noalign{\smallskip}
\multicolumn{9}{l}{\parbox{0.95\textwidth}{\scriptsize
    (a) vdB~86; (b) GU CMa, HRW~2, vdB~88; (c) HRW~3; (d) HRW~4, vdB~89;
    (e) HT CMa, LkHa~218; (f) FZ CMa, HRW~5, vdB~90a, S~295; (g) HRW~6;
    (h) HRW~7; (i) HRW~8, vdB~91;  (j) ALS137;  (k) Z~CMa, HRW~9;  (l)
    HRW~12, vdB~92c;  (m) HRW~13, vdB~92b;  (n) HRW~11, vdB~92a; (o) HRW
    15;  (p) HU CMa, LkHa 220;  (q) HRW~16;  (r) HRW~17;  (s) HRW~18;  (t)
    HRW~19;  (u) IC~2177,  HRW~20, S~292, L~1657;  (v) HRW~23;  (w)
    HRW~24;  (x) HRW~25;  (y) HRW~26;  (z) HRW~27, vdB~94;  (aa)  HRW~28;
    (bb) FM~CMa;  (cc) FN~CMa,  HRW~29, vdB~95; }}\\
\tableline
\endfoot

\tableline
\noalign{\smallskip}
\multicolumn{9}{l}{\parbox{0.95\textwidth}{\scriptsize
    (a) vdB~86; (b) GU CMa, HRW~2, vdB~88; (c) HRW~3; (d) HRW~4, vdB~89;
    (e) HT CMa, LkHa~218; (f) FZ CMa, HRW~5, vdB~90a, S~295; (g) HRW~6;
    (h) HRW~7; (i) HRW~8, vdB~91;  (j) ALS137;  (k) Z~CMa, HRW~9;  (l)
    HRW~12, vdB~92c;  (m) HRW~13, vdB~92b;  (n) HRW~11, vdB~92a; (o) HRW
    15;  (p) HU CMa, LkHa 220;  (q) HRW~16;  (r) HRW~17;  (s) HRW~18;  (t)
    HRW~19;  (u) IC~2177,  HRW~20, S~292, L~1657;  (v) HRW~23;  (w)
    HRW~24;  (x) HRW~25;  (y) HRW~26;  (z) HRW~27, vdB~94;  (aa)  HRW~28;
    (bb) FM~CMa;  (cc) FN~CMa,  HRW~29, vdB~95; }}\\
\tableline
\tableline
\endlastfoot

13	&		&	51361	&	225.15, -4.81	&	6 56 30.3 	&	-13 03 33	&	10.1	&	B9Ib	&\\
18	&		&	51454	&	223.32, -3.73	&	6 57 04.1 	&	-10 56 50	&	9.4	&	B8IV	&\\
20	&		&	51479	&	222.74, -3.42	&	6 57 07.9	&	-10 16 47	&	8.4	&	B7V	&a\\
24	&		&	51542	&	223.51, -3.75	&	6 57 21.2 	&	-11 07 03 	&	9.6	&	B3V	&\\
32	&		&	51785	&	222.00, -2.69	&	6 58 23.6 	&	-09 17 44	&	9.5	&	B6V	&\\
39	&		&	51961	&	223.48, -3.28	&	6 59 00.4 	&	-10 52 55	&	9.6	&		&\\
46	&		&	52112	&	222.32, -2.53	&	6 59 35.3 	&	-09 30 04	&	8.8	&	B3V	&\\
48	&		&	52159	&	223.81, -3.25	&	6 59 42.6	&	-11 09 26	&	9.6	&	B5Vne	&\\
	&	7	&		&	223.58, -2.96	&	7 00 20.6	&	-10 49 14	&	11.4	&	A8-F0	&\\
	&	11	&		&	223.64, -2.95	&	7 00 28.7	&	-10 52 12 	&	10.2	&	B9	&\\
	&	9	&		&	224.39, -3.32	&	7 00 32.1	&	-11 42 26	&	11.9	&	B6	&\\
	&	19	&		&	223.64, -2.91	&	7 00 38.0 	&	-10 51 27	&	11.3	&	B7-8	&\\
	&	18	&	52412	&	224.24, -3.22	&	7 00 38.5	&	-11 31 45	&	10.2	&	B4-6	&\\
	&	16	&		&	224.85, -3.53	&	7 00 38.5	&	-12 12 39	&	11.3	&	A0	&\\
	&	20	&		&	223.66, -2.92	&	7 00 38.6	&	-10 52 34	&	11.6	&	B9	&\\
	&	21	&		&	223.40, -2.78	&	7 00 38.8 	&	-10 34 45	&	11.6	&	B8-9	&\\
	&	23	&		&	223.74, -2.93	&	7 00 44.7	&	-10 57 24	&	11.3	&	B5-6	&\\
	&	25	&		&	224.18, -3.05	&	7 01 07.8 	&	-11 24 05	&	10.8	&	A1-2	&\\
	&	28	&		&	223.89, -2.87	&	7 01 14.9 	&	-11 03 48	&	10.0	&	B3	&\\
	&	30	&		&	223.67, -2.72	&	7 01 22.9 	&	-10 47 42	&	10.9	&	B8	&\\
63	&	158	&	52721	&	224.17, -2.85	&	7 01 49.5 	&	-11 18 03	&	6.6	&	B2Vne	&b\\
	&	38	&		&	224.21, -2.87	&	7 01 49.6 	&	-11 20 39	&	11.7	&               &c\\
	&	40	&	52746	&	224.97, -3.23	&	7 01 58.0	&	-12 10 51	&	9.75	&	B6	&\\
	&	45	&		&	223.26, -2.32	&	7 02 04.9	&	-10 14 37	&	11.0	&	B9	&\\
	&	50	&		&	223.29, -2.30	&	7 02 11.1 	&	-10 15 49	&	10.5	&	B3	&\\
	&	52	&		&	224.70, -2.97	&	7 02 23.7	&	-11 49 40	&	11.1	&	B8-9	&\\
	&	58	&		&	223.72, -2.43	&	7 02 31.4	&	-10 42 43	&	10.2	&	B9	&\\
	&	57	&		&	224.80, -2.95	&	7 02 37.9	&	-11 54 03	&	11.1	&	B7-8	&\\
	&	56	&	-121748	&	225.09, -3.11	&	7 02 38.4	&	-12 13 59	&	10.3	&	B8	&d\\
	&	160	&		&	224.39, -2.73	&	7 02 42.3	&	-11 26 10	&	11.9	&	Ae	&e\\
67	&	159	&	52942	&	224.41, -2.73	&	7 02 42.6 	&	-11 27 11	&	8.1	&	B2.5IV	&f\\
	&	66	&	53011	&	224.89, -2.91	&	7 02 57.7	&	-11 58 10	&	10.1	&	B7-9	&\\
70	&	68	&		&	224.44, -2.68	&	7 02 58.4	&	-11 27 24	&	11.2	&	B2	&g\\
71	&	70	&	53010	&	224.14, -2.51	&	7 03 02.3 	&	-11 07 01	&	8.9	&	B2.5V	&\\
	&	71	&		&	224.62, -2.73	&	7 03 06.8	&	-11 38 29	&	13.1	&		&\\
72	&	72	&	53035	&	224.23, -2.52	&	7 03 07.8	&	-11 11 57	&	7.9	&	B1	&\\
	&	73	&		&	224.63, -2.72	&	7 03 10.6	&	-11 38 27	&	10.6	&	B3	&h\\
	&	77	&		&	224.42, -2.57	&	7 03 19.4 	&	-11 23 27	&	11.7	&	B3	&\\
	&	78	&		&	224.01, -2.36	&	7 03 20.5	&	-10 55 53	&	10.9	&	B2	&\\
75	&	79	&	-101839	&	223.82, -2.24	&	7 03 24.6 	&	-10 42 16	&	9.7	&	B3V &i\\
	&	80	&		&	224.77, -2.68	&	7 03 34.6	&	-11 44 55	&	10.6	&	B0	&j\\
	&	86	&		&	223.61, -2.06	&	7 03 39.3	&	-10 26 13	&	11.5	&	B9-A0 	&\\
	&	82	&		&	224.58, -2.56	&	7 03 39.3	&	-11 31 51	&	11.6	&	B8	& \\
	&	81	&		&	224.84, -2.70	&	7 03 39.4	&	-11 49 35	&	10.5	&	B9	& \\
	&	84	&		&	224.46, -2.49	&	7 03 40.4	&	-11 23 31	&	11.9	&	A	&\\
76	&	161	&	53179	&	224.54, -2.52	&	7 03 43.2 	&	-11 33 06	&	9.9	&	Bpe 	&k\\
	&	88	&		&	224.55, -2.50	&	7 03 49.6	&	-11 28 23	&	11.6	&	        &\\
77	&	92	&	-111762	&	224.56, -2.48	&	7 03 54.4 	&	-11 28 29	&	10.0	&	B2 	&l\\
	&	93	&		&	224.64, -2.52	&	7 03 54.6	&	-11 33 40	&	10.1	&		&\\
78	&	91	&	-111763	&	224.65, -2.53	&	7 03 54.9 	&	-11 34 34	&	8.9	&	B1.5V	&m\\
79	&	89	&	-111761	&	224.65, -2.52	&7 03 55.0	&	-11 34 30	&	9.3	&	B2	&n\\
14	&	94	&		&	224.65, -2.52	&	7 03 56.4	&	-11 34 39	&	12.1	&		&		\\
	&	95	&		&	224.67, -2.52	&	7 03 58.5	&	-11 35 30	&	11.3	&		&		\\
17	&	97	&		&	224.32, -2.34	&	7 03 59.0	&	-11 11 58	&	10.4	&	B5	&o\\
	&	101	&		&	224.15, -2.24	&	7 04 00.7	&	-11 00 03	&	11.2	&	B6-8	&\\
	&	99	&		&	224.53, -2.43	&	7 04 02.3 	&	-11 25 39	&	10.5	&	B3-5	&\\
	&	100	&		&	224.63, -2.48	&	7 04 02.6	&	-11 32 15	&	11.7	&	B8-9	&\\
	&	102	&		&	224.54, -2.43	&	7 04 03.9	&	-11 26 10	&	9.93	&		&\\
	&	103	&	-11 1765&	224.59, -2.44	&	7 04 05.4	&	-11 28 56	&	10.3	&	B2	&\\
	&	162	&		&	224.55, -2.42	&7 04 06.8	&	-11 26 09	&	11.8	&	B5e	&p\\
81	&	107	&		&	224.44, -2.36	&	7 04 06.9	&	-11 18 49	&	11.0	&	B6	&q\\
	&	108	&		&	225.21, -2.73	&	7 04 12.5 	&	-12 10 19	&	11.1	&	B6-8	&\\
	&	111	&		&	224.45, -2.34	&	7 04 13.0	&	-11 19 01	&	11.8	&	B8	&r\\
84	&	114	&	53339	&	224.53, -2.37	&	7 04 15.9 	&	-11 24 05	&	9.4	&	B3V	&\\
	&	112	&	53396	&	224.82, -2.51	&	7 04 17.4 	&	-11 43 10	&	10.7	&	B9	&\\
	&	116	&		&	224.45, -2.32	&	7 04 17.4	&	-11 18 09	&	12.0	&	A1	&s\\
	&	117	&		&	224.44, -2.31	&	7 04 19.0	&	-11 17 13	&	12.1	&	A0	&t\\
	&	119	&		&	224.77, -2.46	&	7 04 21.7	&	-11 39 05	&	10.7	&	B5	&\\
86	&	163	&	53367	&	223.71, -1.90	&	7 04 25.5 	&	-10 27 16	&	7	&	B0IVe	&u\\
	&	121	&		&	224.68, -2.39	&	7 04 26.1	&	-11 32 20 	&	12.3	&	B8	&\\
	&	123	&		&	224.71, -2.39	&	7 04 30.6 	&	-11 33 52	&	11.8	&	B5-7	&\\
	&	124	&		&	224.88, -2.46	&	7 04 34.9	&	-11 44 57	&	10.5	&	B7	&\\
92	&	164	&	53456	&	224.68, -2.34	&	7 04 38.3 	&	-11 31 27	&	7.3	&	B0V	&\\
90	&	125	&	53457	&	225.10, -2.55	&	7 04 38.5	&	-11 59 08	&	9.3	&	B6-7	&\\
	&	127	&		&	225.48, -2.73	&	7 04 44.1 	&	-12 24 32	&	10.8	&	B6	&\\
	&	130	&		&	223.85, -1.88	&	7 04 45.3	&	-10 34 12	&	10.8	&	B8III/IV&v\\
	&	131	&		&	224.18, -2.04	&	7 04 47.4	&	-10 56 18	&	13.3	&		&w\\
	&	129	&		&	224.33, -2.12	&	7 04 47.5	&	-11 06 37	&	13.2	&		&x\\
	&	132	&		&	224.47, -2.17	&	7 04 53.5	&	-11 15 19	&	11.7	&	B6-7	&\\
	&	135	&		&	223.93, -1.88	&	7 04 54.1 	&	-10 38 50 	&	12.4	&	A1 IV	&y\\
	&	137	&		&	223.70, -1.73	&	7 05 02.1	&	-10 22 04	&	11.4	&	B6-8	&\\
	&	139	&		&	223.71, -1.72	&	7 05 04.5	&	-10 22 33	&	11.9	&	B9-A0	&\\
	&	140	&	53569	&	224.12, -1.92	&	7 05 07.2	&	-10 50 01	&	10.4	&	B9	&\\
	&	141	&		&	224.14, -1.93	&	7 05 07.5	&	-10 51 19	&	9.8	&	B9	&\\
96	&		&	53595	&	224.37, -2.03	&	7 05 11.8	&	-11 06 02	&	9.2	&	B5Vn	&\\

97	&	142	&	53623	&	225.47, -2.57	&	7 05 16.8 	&	-12 19 34	&	8	&	B1 II/III&z\\
98	&	144	&	53622	&	224.55, -2.09	&	7 05 18.5 	&	-11 17 22	&	9.4	&	B5V	&\\
	&	143	&		&	224.84, -2.24	&	7 05 18.8	&	-11 36 49	&	10.5	&	B6-8	&\\
	&	145	&		&	224.48, -2.02	&	7 05 26.4	&	-11 11 22	&	11.5	&	B7-8	&\\
	&	146	&		&	223.70, -1.58	&	7 05 34.0	&	-10 18 17	&	10.9	&	B8-9	&\\
100	&		&	53667	&	222.31, -0.85	&	7 05 35.2 	&	-08 43 44	&	7.8	&	B0.5III	&\\
101	&	149	&	53691	&	224.47, -1.96	&	7 05 37.6 	&	-11 09 26	&	9.4	&	B3V	&aa\\
102	&		&	53756	&	225.95, -2.70	&	7 05 42.1 	&	-12 48 43 	&	7.3	&	B2/B3II	&bb\\
104	&	165	&	53755	&	223.98, -1.65	&	7 05 49.6 	&	-10 34 57 	&	6.5	&	B0.5V	&\\
	&	156	&	-10 1866&	224.18, -1.73	&	7 05 53.9	&	-10 47 43	&	10.4	&	B4-6	&\\
114	&		&	53948	&	225.94, -2.46	&	7 06 33.6	&	-12 41 53	&	9.8	&	B5	&\\
115	&		&	53975	&	225.68, -2.32	&	7 06 36.0 	&	-12 23 38	&	6.5	&	B7Iab/Ib&\\
116	&		&	53974	&	224.71, -1.79	&	7 06 40.8 	&	-11 17 38 	&	5.4	&	B0.5IV	&cc\\
120	&		&	54025	&	224.77, -1.75	&	7 06 56.4 	&	-11 19 38	&	7.7	&	B1V	&\\
131	&		&	54306	&	225.40, -1.82	&	7 07 52.9 	&	-11 54 50	&	8.7	&	B2V	&\\
133	&		&	54439	&	225.40, -1.68	&	7 08 23.2 	&	-11 51 09	&	7.2	&	B2III	&\\
134	&		&	54493	&	226.33, -2.12	&	7 08 32.2 	&	-12 53 09	&	7.2	&	B2IV	&\\
145	&		&	54672	&	240.45, -9.21	&	7 08 33.7 	&	-28 36 47	&	7.5	&	F7V	&\\
138	&		&	54574	&	225.14, -1.41	&	7 08 53.9 	&	-11 30 05 	&	8.8	&	B2V	&\\
141	&		&	54662	&	224.17, -0.78	&	7 09 20.3	&	-10 20 48	&	6.2	&	O7III	&\\
150	&		&	54879	&	225.55, -1.28	&	7 10 08.1	&	-11 48 10	&	7.7	&	O9.5V	&\\
160	&		&	55096	&	232.97, -4.94	&	7 10 44.5 	&	-20 03 56 	&	9.5	&	A1IV/V	&\\
161	&		&	55121	&	226.30, -1.39	&	7 11 08.0 	&	-12 31 22	&	8.6	&	B3II/III&\\
167	&		&	55266	&	224.96, -0.51	&	7 11 46.5 	&	-10 55 21	&	10	&	B5V	&\\
177	&		&	55442	&	226.16, -0.95	&	7 12 29.3	&	-12 11 42	&	9.5	&	B3V	&\\
178	&		&	55439	&	224.09, 0.16	&	7 12 34.8 	&	-09 50 42	&	8.6	&	B2Ve	&\\
184	&	 	&	55562	&	226.58, -1.03	&	7 12 58.1 	&	-12 36 21 	&	9.0	&	B3V	&\\
190	&		&	55776	&	224.53, 0.33	&	7 14 01.2	&	-10 09 03	&	10.0	&	B8V	&\\
\end{longtable}
} 

\end{center}


\begin{figure}[!htbp]
\centering
\includegraphics[draft=False,width=0.7\textwidth]{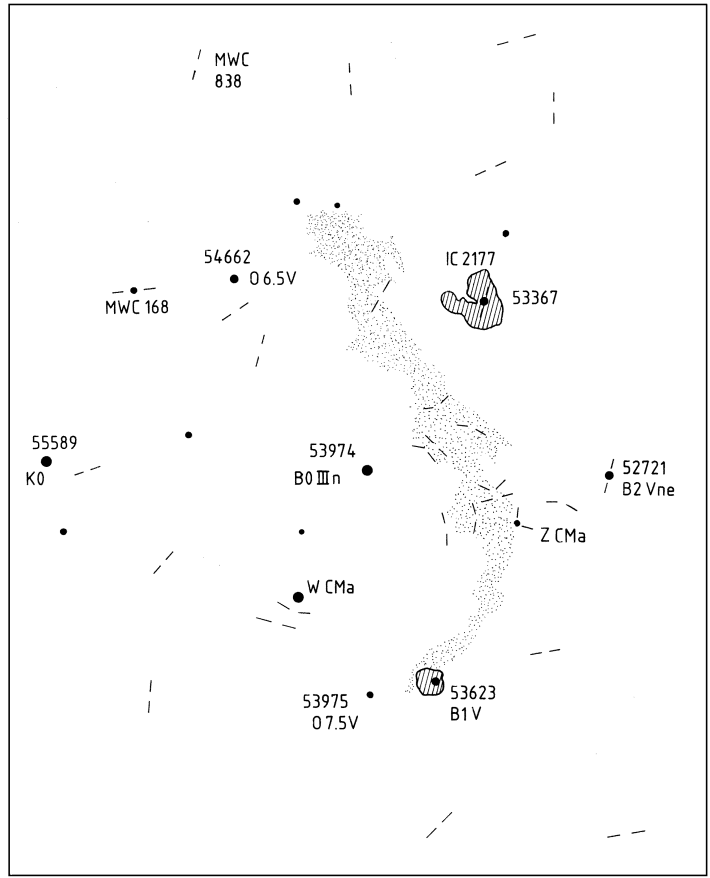}
\caption{Schema extracted from Herbig (1991) showing the
identification of early-type stars (by their HD numbers) and
reflection nebulae. The marks show the positions of emission-line stars
found by Wiramihardja et al. (1986). The field covers 4.1 $\times$ 3.3 square degrees. North is up and East is to the left.}
\end{figure}

\begin{figure}[!htbp]
\centering
\vspace{-5cm}
\includegraphics[draft=False,width=1.0\textwidth]{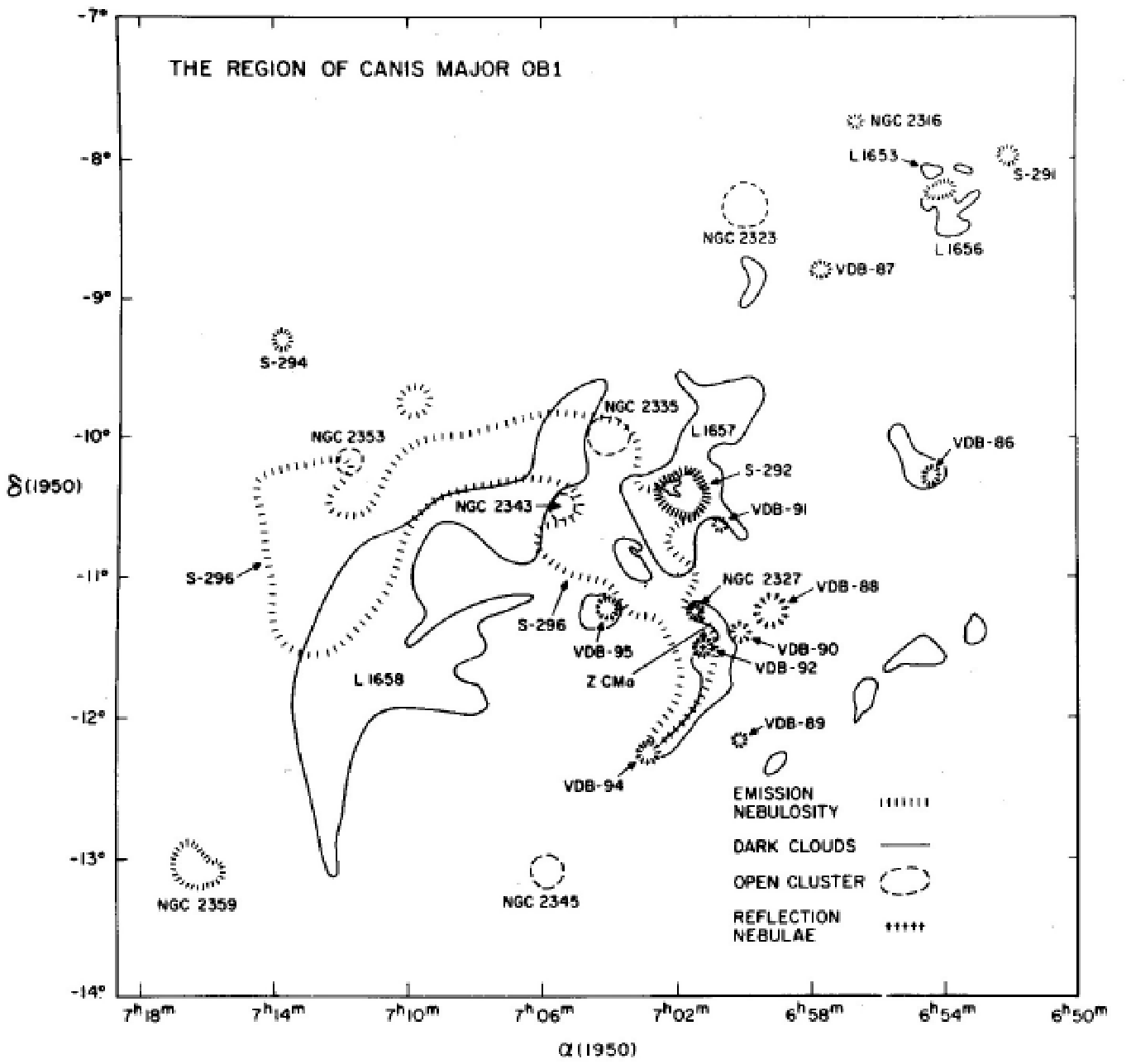}
\vspace{-5cm}
\caption{The distribution of the conspicuous objects found in CMa OB1/R1 region.
Two dark clouds, L1658 and L1657, correspond to the main structures detected
in the map of integrated $^{12}$CO emission obtained by Blitz (1980). This
drawing is extracted from Fig. 8 in the paper by Blitz (1980). }
\end{figure}

Eggen (1981) proposed that stars in the Cr~121 region form two
groups. The nearer concentration of stars located mainly within a
$2^o$ circle centered on $(\alpha,\delta)_{J2000}$ = ($6^h 54^m, -23^o
37.5'$) is probably related to NGC~2287, because they have similar age
and motion. He suggested that this is a separate association, CMa OB2.

The similarities in age and distance of Cr~121 and CMa OB1 points to a
physical relation leading Eggen to argue that the Cr~121 cluster is
probably an extension of CMa OB1.  On the other hand, de Zeeuw et
al. (1999) used {\it Hipparcos} data to identify Cr~121 as a moving
group not related with CMa OB1. The controversy on the nature of Cr~121
is discussed further in Sect.~4.

CMa~R1 contains several OB stars associated with the ring of emission
nebulae coinciding with an expanding HI shell that is suggested to be
a supernova remnant (SNR) inducing the star formation in this region
(Herbst \& Assousa 1977).  Linear polarization observations are
consistent with a model of compression by a supernova shock (Vrba,
Baierlein, \& Herbst 1987) but stellar winds could also produce a
similar structure and morphology.  A more detailed discussion on the
hypothesis of star formation in CMa R1 induced by SNR is presented in
Sect.~7.

\section{Cloud Structure}

\subsection{Dust Distribution and Visual Extinction}

\begin{figure}[tbp]
\vspace{-2.0cm}
\centering
\includegraphics[draft=False,width=01.0\textwidth]{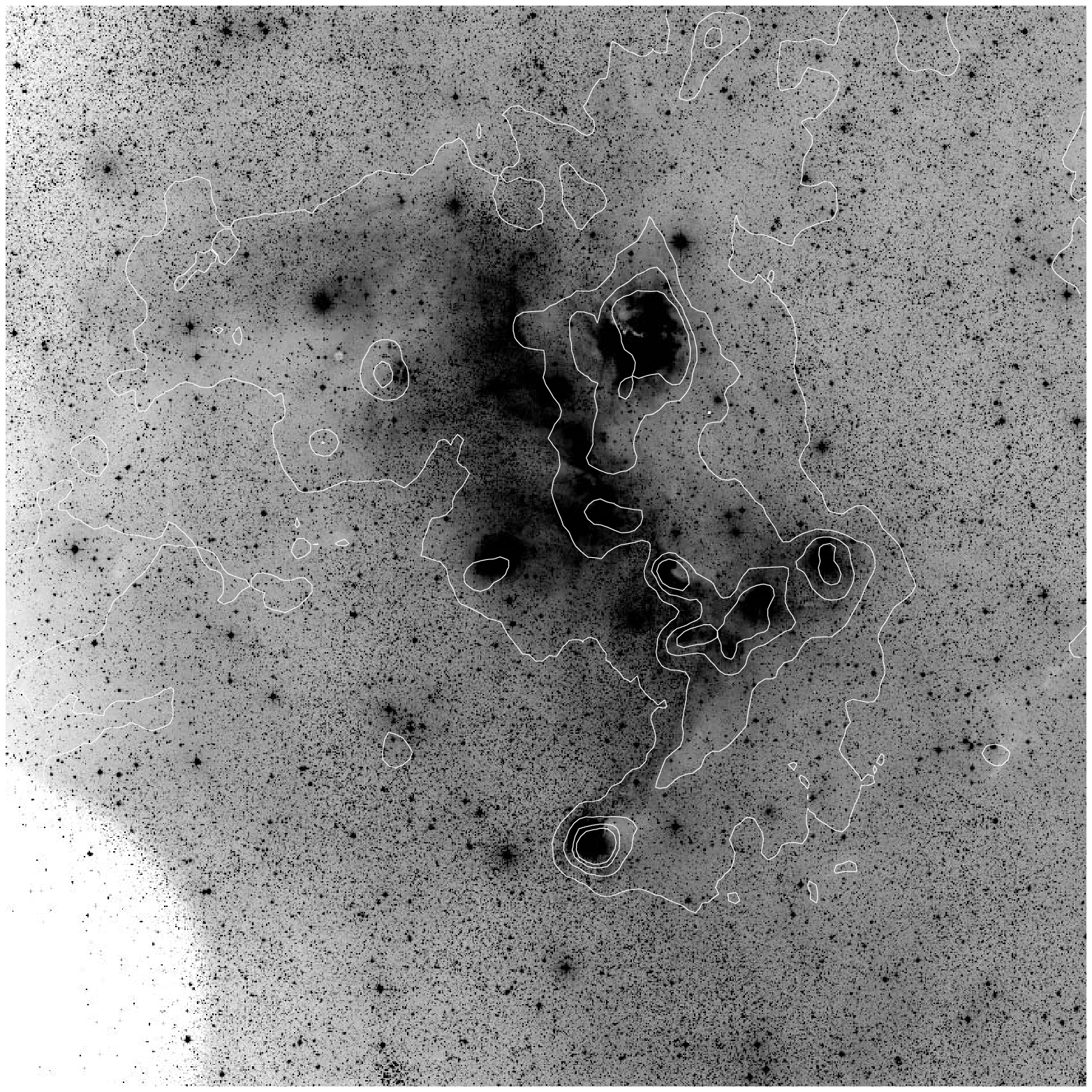}
\vspace{-5.0cm}
\caption{Far-infrared contours superimposed on the optical image of the CMa
OB1/R1 region obtained from the Digitized Sky Survey - DSS 2 plate 170. The contours,
showing levels 32, 80, 160, and 240 MJy/sr, correspond to  IRAS-ISIS data obtained at
100 ~$\mu$m. Size and orientation are the same as Figure 2.}
\end{figure}

A picture of the dust distribution of the Canis Majoris cloud complex
is given in Figure~5, which shows the IRAS contours at 100 ~$\mu$m
superposed on a DSS optical image. The higher levels of far-infrared emission
are associated to emission or reflection nebulae, mainly around
 the stars HD~53367 (IC~2177), GU~CMa, Z~CMa, and HD~53623. The overall
contours show the correlation of gas and dust distributions.

\begin{sidewaysfigure}[p]
\centering
\includegraphics[draft=False,height=11cm]{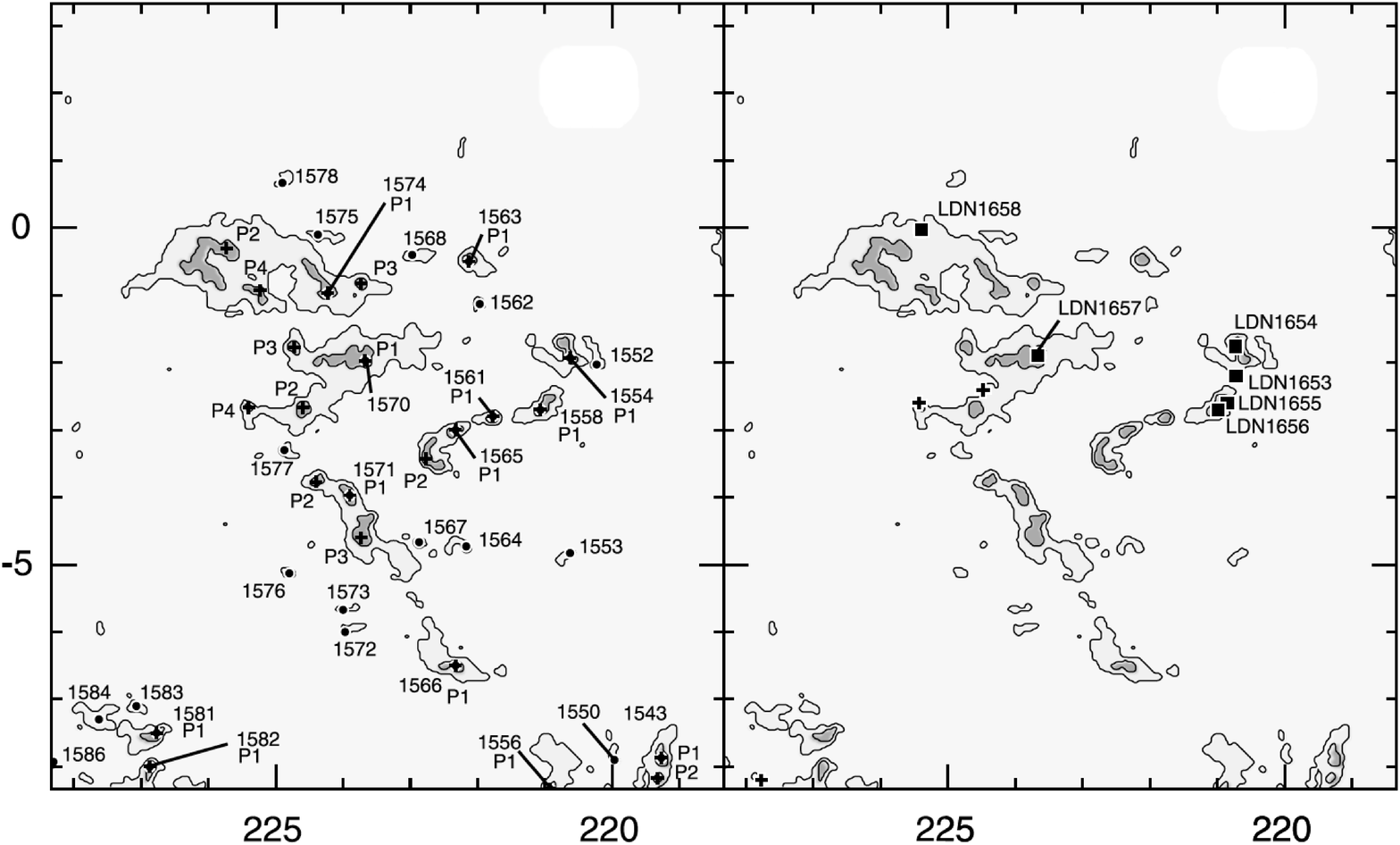}
\caption{The spatial distribution of the CMa clouds extracted from Dobashi et al. (2005), presented in Galactic
coordinates  (latitude {\it versus} longitude). The left panel shows
Dobashi et al.'s cloud identifications (TGU). High extinction cores are labeled with ``P". The right side shows the nominal
positions from Lynds catalog.}
\end{sidewaysfigure}

Figure~6 presents the CMa cloud distribution extracted from the atlas
of dark clouds by Dobashi et al. (2005) and shows the identification
of Lynds clouds, Dobashi clouds (named TGU), and high extinction cores
(labeled P).

An extinction map of the CMa R1 region based on star counts using the J band data from
the near-infrared (near-IR) {\it DENIS} Catalogue has been performed by Cambr\'esy (2000).
 In the region towards the dark
clouds the estimated $A_J$ levels are $\sim$ 1 to 4 mag.  The maximum
extinction provided by the star counting method is a lower limit of
$A_J >$ 6.8 mag. Adopting the relation between extinctions in optical
and infrared bands given by Cardelli et al. (1989), a corresponding
visual extinction of $A_V >$ 24 mag can be inferred.

Herbst, Racine, \& Warner (1978) obtained optical and near-IR
photometry as well as spectral types for 30 stars illuminating
reflection nebulae in CMa~R1.  Several of these stars present large
V-K colors suggesting a steeper than normal extinction law. Later,
Herbst et al. (1982) examined the reddening law applicable to some of
these reflection nebulae and verified that the ratio of total to
selective extinction is about 4.2. Based on the detection of very
large K-L excesses, the presence of circumstellar shells is indicated
for two stars (CMa~R1-16 and CMa~R1-18 in Herbst et al.'s Table~I).
A similar anomalous extinction was found by Vrba et al. (1987) based
upon the wavelength dependence of polarization of stars within CMa R1.
The interstellar extinction law towards CMa~R1 was also studied by
Terranegra et al. (1994). They used {\it uvby} photometry to derive
$R_y=5.4 ~(R_v=4.0)$ similar to the value estimated by Herbst et
al. (1982).

\subsection{Gas Distribution}

Blitz (1980) has presented a CO map with 10$'$ angular resolution that
revealed a molecular emission arising from two distinct clouds. The
most intense emission appears to be associated with the HII region
S292, and the second cloud seems to be related to S296. A length of
101~pc and a mass of $3\times 10^4 M_{\odot}$ were derived for the
cloud complex in CMa OB1, based on CO data and adopting a distance of
1.1~kpc.

A more detailed mapping of the region in CO is presented by Machnik et
al.  (1980), who found 23 peaks of CO emission. A reproduction of this
map is shown in Figure~7.

\begin{figure}[!tb]
\centering
\vspace{-11cm}
\includegraphics[width=1.3\textwidth,draft=False]{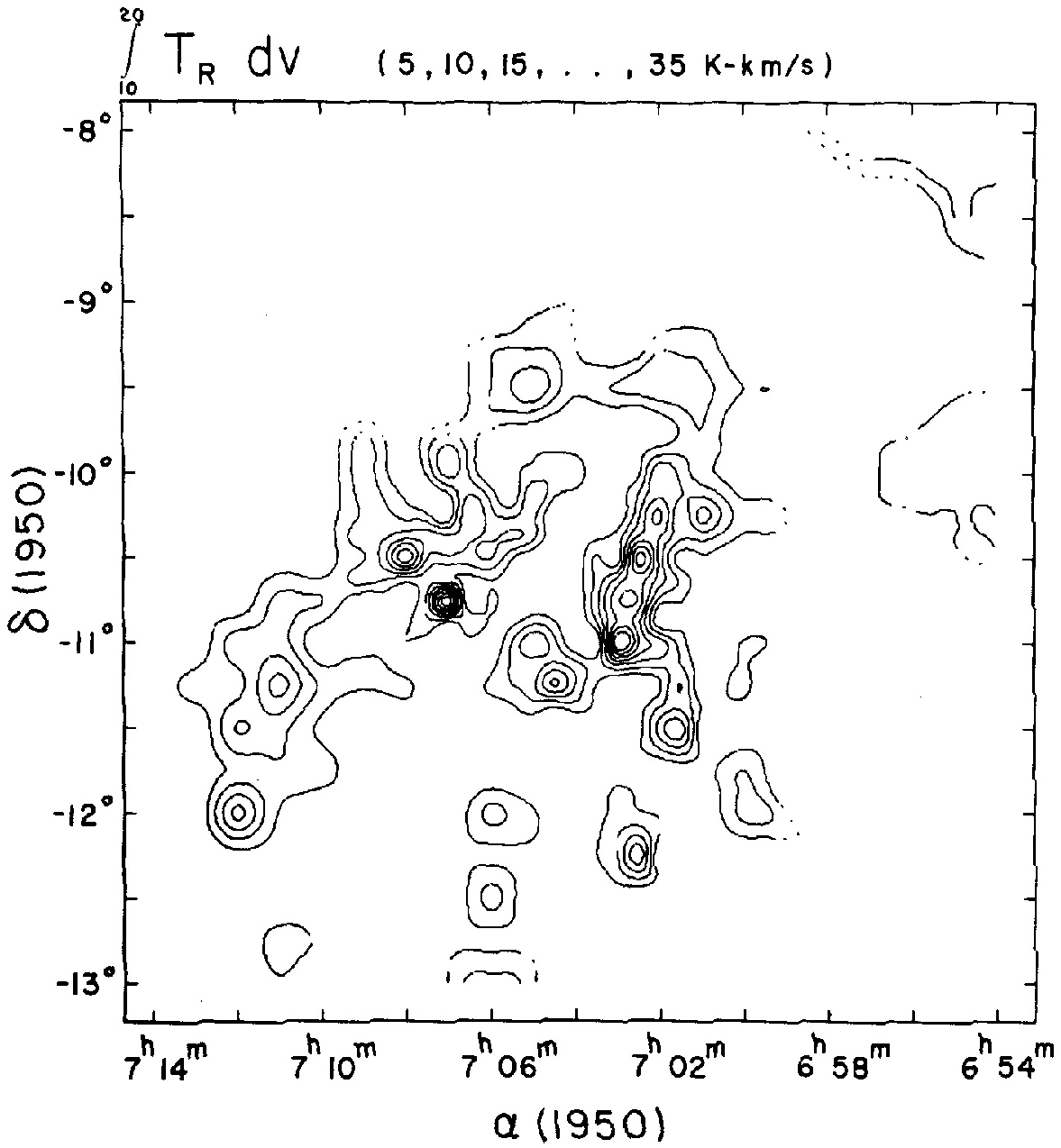}
\caption{CO map of the central region of CMa OB1, obtained at 2.6~mm. Extracted from Machnik et al. (1980).}
\end{figure}

Kim et al. (2004) developed a large scale $^{13}$CO survey covering
two groups in the CMa~OB1 and G~220.8-1.7 (L1656) regions, where they
found 22 clouds.  The mass spectrum of the clouds shows a power-law
with index $-1.55\pm0.09$. A comparative study revealed that the
fraction of star-forming clouds is higher among the massive clouds ($M_{cloud}
> 10^{3.5}M_{\odot}$).  Strong UV radiation from O-type stars in the
vicinity of the clouds may explain the small number of low-mass clouds
in CMa~OB1. Some of the results derived from this survey are presented
in Table 2, which lists various objects associated with the $^{13}$CO
clouds. Figure~8 shows the distribution of the clouds compared with
the position of identified objects, as presented by Kim et al. (2004).

\begin{sidewaystable}[tbp]
\caption{List of $^{13}$CO clouds observed by Kim et al. (2004) towards the CMa
region. Cloud name, Galactic coordinates, area and mass of the clouds are given.
Last column is used to list the catalogued dark clouds, H II regions, and
reflection nebulae.}
\smallskip\begin{center}
{\small
\begin{tabular*}{8.3in}{@{\extracolsep\fill}rcccccl}
\tableline
\noalign{\smallskip}
 No.& Cloud  & {\em l} & {\em b} & Area & $M_{cloud}$ & Objects associated with the
$^{13}$CO
clouds\\
    & name   &($^o$) &($^o$) & $(pc^2)$ &   ($M_{\odot})$  & \\
\noalign{\smallskip}
\tableline
\noalign{\smallskip}
1 & 222.8-00.4 & 222.80 & -0.40 & 50 & 1250 &  \\
2 & 223.5-00.9 & 223.47 & -0.93 & 29 & 620 & S296 \\
3 & 223.9-01.9 & 223.87 & -1.87 & 358 & 16000 & L1657, TDS652, S292, S296,
vdB93,
vdB95\\
4 & 224.3-01.1 & 224.27 & -1.07 & 301 & 12000 & L1658, TDS653, S296\\
5 & 224.4-02.4 & 224.40 & -2.40 & 43 & 700 & TDS654, S296\\
6 & 224.4-02.8 & 224.40 & -2.80 & 14 & 230 & S295, S296, vdB90\\
7 & 224.7-02.5 & 224.67 & -2.53 & 43 & 890 & S296, vdB92\\
8 & 224.9+00.8 & 224.93 & 0.80 & 57 & 640 & \\
9 & 224.9-01.2 & 224.93 & -1.20 & 29 & 720 & S296\\
10 & 225.3-01.1 & 225.33 & -1.07 & 43 & 600 & S296 \\
11 & 225.5-02.8 & 225.57 & -2.80 & 43 & 600 & S297, S296, vdB94\\
12 & 226.1-00.4 & 226.13 & -0.40 & 315 & 7500 & L1658, S296\\
13 & 226.5+01.3 & 226.53 & 1.33 & 57 & 860  & \\
14 & 219.7-02.4 & 219.73 & -2.40 & ? & ? & \\
15 & 220.4-02.3 & 220.40 & -2.27 & 22 & 600 &  \\
16 & 220.8-01.7 & 220.80 & -1.73 & 87 & 2200 & L1654, BFS62 \\
17 & 220.9-02.7 & 220.93 & -2.67 & 70 & 1700 & L1653, L1655, L1656, BFS61,
BFS63\\
18 & 221.2-02.4 & 221.20 & -2.40 & 43 & 1100 & BFS63 \\
19 & 221.5-01.3 & 221.47 & -1.33 & 38 & 760 & \\
20 & 221.6-02.4 & 221.60 & -2.40 & 27 & 470 & \\
21 & 221.7-02.8 & 221.73 & -2.80 & 32 & 530 & \\
22 & 222.4-03.1 & 222.40 & -3.07 & 22 & 250 & \\
\noalign{\smallskip}
\tableline
\end{tabular*}
}
\end{center}
Notes: Clouds Nos. 1 to 13 are located in the CMa OB1 region, and
 Nos. 14 to 22 are in the G220.8-1.17 region. The related objects are
listed in the following catalogues: dark clouds L (Lynds 1962), and TDS
(Taylor et al. 1987); H II regions S (Sharpless 1959), and BSF (Blitz et al.
1982); reflection nebulae vdB (van den Bergh 1966).
\end{sidewaystable}

Optical absorption line observations of the interstellar $CH^+$
towards CMa OB1 were reported by Gredel (1997), who found a
correlation of $N(CH^+)$ with the optical depth of the clouds. The
results suggest that dissipation of interstellar turbulence is probably
the main mechanism of production of $CH^+$. Gredel reported $CH^+$ and
$CH$ measurements towards the stars HD~55879, HD~53975, HD~53755, HD~54662,
and HD~52382.

A faint, large-scale H$\alpha$ filament, 80$^o$ long and 2$^o$ wide,
that extends perpendicular to the Galactic plane at {\em l}~=~225$^o$ was
discovered by Haffner, Reynolds, \& Tufte (1998). This feature appears
to meet the Galactic plane above CMa OB1 at {\em l} $\sim$225$^o$,
{\em b}~$\sim$0$^o$. The radial velocity varies from v$_{LSR}$ = +16 ~km/s to $-$20
~km/s. The feature seems to have no correspondence to other structures
revealed by different wavelength surveys, and the source of ionization
was not identified.

\begin{figure}[!ht]
\plotone{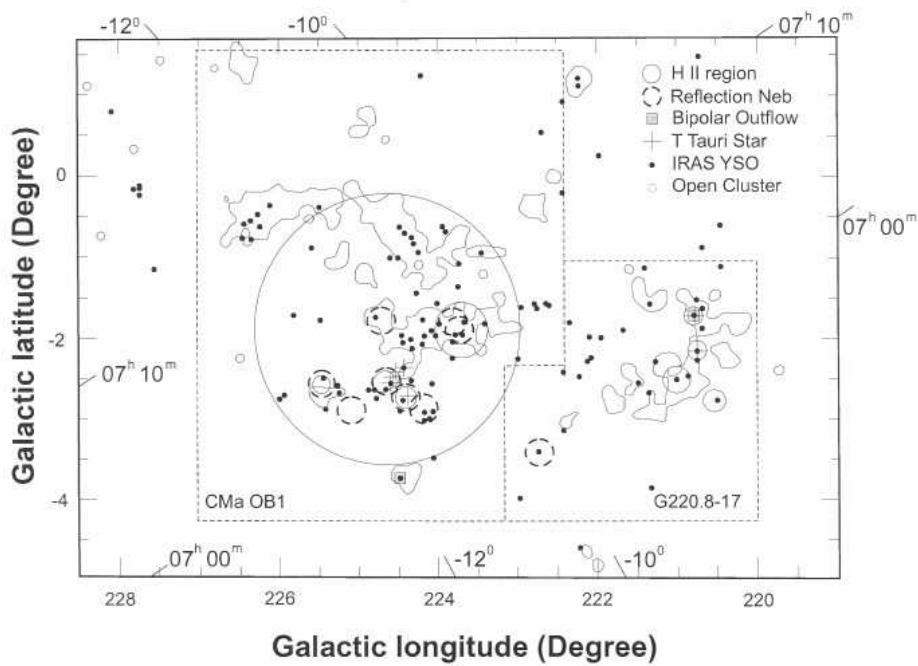}
\caption{Distribution of CO clouds adapted from Kim et al. (2004). Dotted lines indicate
the  CMa OB1 and G220.8-1.7 cloud complexes. The boundary of each cloud is
indicated by the  1~K ~km ~s$^{-1}$ level from the $^{13}$CO contour map.
Black dots mark YSOs detected by IRAS.
A few tick marks are inserted around
the margins to indicate the equatorial J2000 coordinates related to the other figures.}
\end{figure}

\subsection{Infrared Emission}

The distribution of the mid-IR emission in the CMa OB1/R1 region
is shown in Figure~9, an image obtained in the 8.3~$\mu$m band from
the MSX data base. The diffuse emission has a distribution similar to the 100~$\mu$m map,
but more detailed filamentary structures are shown near ({\em l, b}) = (224.2, --2.8) and (223.8, --2),
for example, corresponding to IC~2177 and GU~CMa, respectively.

\begin{figure}[!ht]
\plotone{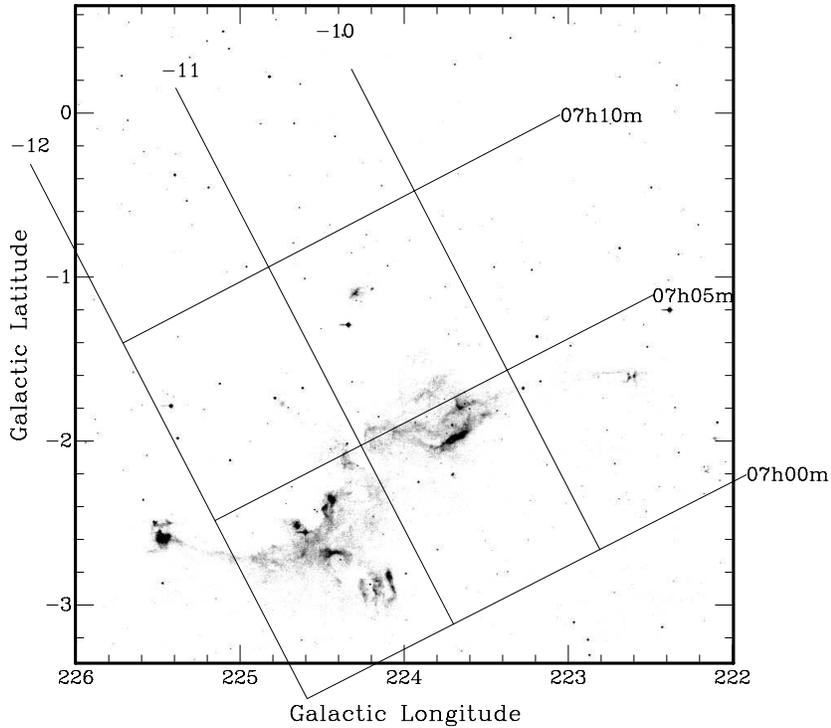}
\vspace{-2cm}
\caption{Mid-infrared map of the Canis Major region, obtained by MSX at
8.3 ~$\mu$m.}
\end{figure}

Luo (1991) used IRAS data to study the infrared emission of the
complex CMa OB1/R1 and to compare it to optical, CO and radio
continuum observations.  The results suggest a composition of diffuse
emission and several discrete sources. Two O stars are probably
responsible for the diffuse emission which, in combination with the
remnant of an old HII region, could produce the extended HII
region. Most of the discrete sources have optical counterparts and
correspond to the emission or reflection nebulae. The sources which do
not have optical counterparts may be excited by early-type stars
that are very
young and still embedded in dense dust clouds.

The far-IR excess was used as a criterion by Kim et al. (2004) to
identify 115 young stellar object candidates, selected
from the IRAS point source catalog.

\section{Stellar Content}
\vspace{-4mm}
Six massive stars are associated with the nebulae around S296. One of
them is Z~CMa, a young luminous star partly embedded in a dark cloud
at the outer edge of S296 (Herbig 1960). Hartmann et al. (1989)
classified it as a FU Orionis (FUor) type star. Later, it was
discovered that Z~CMa is a double system. The primary component is a
Herbig Be star and the secondary is a FUor object
(Koresko et al. 1991). More details about Z CMa are given in
Sect. 3.2. Another interesting object is the unusually bright,
variable carbon star W~CMa. Assuming that it is a member of the
association, and adopting the reddening from the neighboring stars,
Eggen (1978) estimated an extreme luminosity for W~CMa,
$M_{bol} \sim -7$ mag, similar to the bright carbon stars in the Large
Magellanic Cloud.
\vspace{4mm}

\subsection{Emission Line Stars}

Photometric data for large samples of stars located in the CMa OB1
region have been reported by several authors: Feinstein (1967), Eggen
(1981), de Geus et al. (1990), and Shevchenko et al. (199).  Herbig \&
Rao (1974) first identified the emission line stars LkH$\alpha$~218
(HT~CMa) and LkH$\alpha$~220 (HU~CMa).  A survey of H$\alpha$ emission
line stars made by Wiramihardja et al. (1986) in a 6$^\circ$ $\times$
7$^\circ$ area around CMa OB1 revealed 179 emission line stars, for
which {\it UBV} photometry was carried out.

The detection of new B stars associated with the CMa~R1 star forming
region was reported by Terranegra et al. (1994). Based on
color-magnitude diagrams, four stars were found to be probable new
members associated with CMa~R1.

Schwartz, Persson, \& Hamann (1990) reported the identification of emission-line
objects in southern regions. In Canis Major, their study covered a $\sim$ 2$^o$ field
around $\alpha=7^h 25^m$, $\delta=-26^o$ (J2000), where 26 emission line stars were identified.
Based upon spectroscopic data obtained for these objects, Pereira et al. (2001) proposed
that 16 of them are Be stars and seven are T Tauri stars.

A more detailed study of the stellar composition of CMa~R1 was
presented by Shevchenko et al. (1999) in photometric and spectroscopic
surveys of 165 stars brighter than V=13~mag, revealing that 88 of them
are members of the association. Most of the members are considered
pre-main sequence (PMS) stars with ages $<$ 6 $10^6$ yr. Two bright B stars
GU~CMa and FZ~CMa seem to be older than the association and are
suggested not to be formed in the same star formation episode. Nine of
the CMa~R1 members showing circumstellar dust were studied by
Tjin A~Djie et al. (2001). They suggested that four stars of the
sample are Herbig Be stars with disks.  The variability of these stars
is interpreted in terms of: {\it(i)} eclipsing binary (HD~52721); {\it(ii)}
magnetic activity cycle (HD~53367); or {\it(iii)} circumstellar dust
variations (LkH$\alpha$~220 and LkH$\alpha$~218). The other five stars
of the sample, located inside the arc-shaped border of the HII
region, do not show indications of circumstellar disks.  The absence
of a circumstellar disk can be explained by the possible evaporation
by UV photons from nearby O stars, or from the nearby supernova that
could have occurred about 1 ~Myr ago.

\subsection{Z CMa}

The young star Z~CMa has been subject to numerous detailed
studies. Different spectroscopic and photometric data have been
reported: optical spectra obtained by Covino et al. (1984),
Finkenzeller \& Jankovics (1984), and Finkenzeller \& Mundt (1984);
ultraviolet spectra by Kenyon et al. (1989); photometric variability
investigated by Kolotilov (1991). Long term variability in the optical
and near-IR were reported by Covino et al. (1984), Teodorani et
al. (1997) and van den Ancker et al. (2004). Zinnecker \& Preibisch (1994)
detected X-ray emission from the star using {\it ROSAT} observations,
which are detailed in Sect. 6.

Significant spectral changes were discussed by Covino et
al. (1984). The FUor nature of Z~CMa was first suggested by Hartmann
et al. (1989) and confirmed by Hessman et al. (1991) and other authors
(Whitney et al. 1993, Torres et al. 1994). Near-IR speckle
observations by Leinert \& Haas (1987) and Koresko et al. (1989)
revealed an elongated emission, with a size of about 0.1 arcsec at
2.2~$\mu$m. The presence of a companion at PA=$122^o$, with separation
of 0.1 arcsec, was later confirmed using more detailed near-IR speckle
interferometric observations by Koresko et al. (1991) and Haas et
al. (1992).  Large amounts of cold dust around Z~CMa were revealed by
sub-millimeter observations reported by Weintraub et
al. (1991). Emission at centimeter wavelengths was also detected by
Cohen \& Bieging (1986).

A major, well-collimated Herbig-Haro jet, HH~160, emanating from Z CMa
was discovered by Poetzel, Mundt, \& Ray (1989).  The origin of the
jet has been the subject of discussion based on polarization studies
(Whitney et al. 1993, Fisher et al. 1998, Lamzin et al. 1998).
Conflicting results were also obtained using high resolution optical
spectroscopy and high angular resolution radio observations. Garcia et
al. (1999) identified a microjet associated with the infrared
companion, based on speckle interferometry covering the
[OI]$\lambda$6300 line of both components with spatial resolution of
0.24 arcsec.  The [OI] centroid position is blueshifted in the
spectrum of the secondary (the optical object) and redshifted in the
primary (near-infrared object), which was interpreted to imply that
the primary drives the jet. They also used a deconvolved image of the
[OI] emission to show evidence of a microjet whose axis passes through
the primary.  In contrast, Vel\'azquez \& Rodr\'\i guez (2001) used
sensitive, high resolution VLA observations at 6~cm and 3.5~cm to
establish that the jet originates from the optical component. With an
astrometric accuracy of 0.002 arcsec, the VLA contour map at 3.5~cm
shows the origin of the jet associated with a compact radio source,
whose core has an extension to the north-east. Adopting the position
of the optical component given by Perryman et al. (1997) and the
offsets of the infrared component position by Thi\'ebaut et
al. (1995), Vel\'azquez \& Rodr\'\i guez verified that the core of the
radio jet coincides with the optical component. Their conclusions
appear to solve the uncertainty on the origin of the jet, which seems
to be associated with the secondary (optical) component.

Mader (2001) used AAO/UKST H$\alpha$ wide-field material to search for
giant HH flows in CMa OB1/R1 region. Based on these data it is claimed
that the HH flow from Z~CMa seems to be much larger than previously
thought.  Chochol et al. (1998) observed large variations on
H$\alpha$, H$\beta$, and NaI~D P~Cygni absorption detected in
high-resolution, time resolved data.  Several observational facts,
mainly related to the variations of the features in multi-component
structures, are discussed in terms of wind models of FUor stars. The
variations can be explained by equatorial winds and high inclination
of the disk.

A new, short outburst of Z CMa was recently reported by van den Ancker
et al. (2004), based on the observations of a sudden rise in visual
brightness, followed by a rapid decline. This behavior was similar to
another eruption that occurred in 1987.  Evolutionary calculations show
that the system may consist of a 16$M_{\odot}$ B0IIe primary star (infrared
component) and a 3$M_{\odot}$ FUor secondary (optical component), with
a common age of about 3x$10^5$ yrs.
The strength of emission lines originating from the envelope of
the primary star showed strong increase with the brightness, which can
be interpreted in terms of an accretion disk with atmosphere associated
with the Herbig Be component.

Both sources were observed by the Keck interferometer
in the K band.  The FUor object was discussed by Millan-Gabet et al.
(2006), who verified that spatial scales of near-infrared emission
are large compared to the predictions from simple models of accretion
disks. They suggested that disk models require modifications or that
additional model components are required. A definitive analysis of the
data was not possible for the Herbig Ae component  (Monnier et al. 2005).
Sch\"utz, Meeus, \& Sterzik (2005) obtained N-band photometry and
spectroscopy for the FUor component, which showed a broad absorption
band attributed to silicates.
Polomski et al. (2005) published high
spatial resolution mid-IR imaging and spectroscopy, sub-millimeter
photometry, and 3-4~$\mu$m photometry.  A simple radiative transfer
model was adopted to fit the mid-IR observed spectra to derive a dust
temperature of T=274 ~K and the sub-millimeter flux was used to derive a
dust mass of $8 \times 10^{-3} M_{\odot}$.

\section{Clusters}

The relation between CMa OB1 and the open clusters NGC~2353,
~2343
and ~2335
 was discussed by Clari\'a (1974b) based on a comparison of
their distances and ages. The conclusion of this study was that NGC
~2353 seems to be the nucleus of CMa OB1, as proposed by Ruprecht
(1966). The physical connection of the OB association with CMa R1 and
NGC~2353 was also supported by Herbst, Racine \& Warner (1978) and by
Eggen (1978).  However, this was contradicted by the detailed study of
Fitzgerald et al. (1990), see Sect.~1.  Also, NGC~2343 and
~2335 have ages
of about $10^8$ yrs, which indicates they were formed before the
association and are not related to it.

As mentioned in Sect. 1, the relation between the cluster Cr~121 and
CMa~OB1 was discussed by Eggen (1981), suggesting the separation of
two groups of stars in this direction: Cr~121 is considered a compact
open cluster located at the same distance as CMa OB1, and NGC~2287, a
more diffuse group, forms a nearer association, suggested to be CMa
OB2.  More recently, de Zeeuw et al. (1999) studied the structure and
the evolution of nearby young clusters, based on {\it Hipparcos} data.
Astrometric members were identified in young stellar groups out to a
distance of $\sim$ 650 ~pc. A moving group in the CMa OB1/R1 region
could not be identified because the candidate members have small
proper motions and parallaxes. In the direction of Cr~121, de Zeeuw et
al. selected 103 stars, for which a mean distance of 592$\pm$28 ~pc
was derived. The proper motions for 43 stars that were previously
considered Cr~121 members (Collinder 1931; Schmidt-Kaler 1961;
Feinstein 1967; Eggen 1981) were analysed by de Zeeuw et al.  to
reveal evidence for a moving group, but 25 of the 43 stars were
rejected by their selection procedure. Based on the differences of the
mean proper motions, de Zeeuw et al. claim that Eggen's proposed
division in two groups is unphysical. However, the fact that {\it
Hipparcos} parallaxes are unreliable at large distances must be
considered, indicating that these data could only detect the more
diffuse foreground association. The same biased selection method was
used by Dias et al. (2001) to classify as Cr~121 members only the
nearer stars of the field, and to determine a distance of 470 ~pc and
an age of 11 ~Myr for this group of stars (Dias et
al. 2002). Burningham et al. (2003) studied the low mass PMS stars
discovered by {\it XMM-Newton} and {\it ROSAT} observations towards Cr
121. Their results are not in agreement with the scenario proposed by
de Zeeuw et al., in which these low-mass stars are associated with the
nearer moving group suggested to be Cr~121. Burningham et al. argue
that the PMS stars are associated with a young cluster located at 1050
~pc. They confirm that the moving group detected by de Zeeuw et al. is
a foreground association and should be called CMa OB2, in agreement
with Eggen's proposition. Kaltcheva (2000) also suggested the
separation of two groups: Cr~121 is a more compact cluster at 1085
~pc, and the other group of stars lies at 660 - 730 ~pc, having
characteristics of an OB association.  More recent results were
obtained by Kaltcheva \& Makarov (2007) using corrected Hipparcos
parallaxes, which indicated a distance of 750-1000 pc for Cr~121, in
agreement with the previous results based on photometric parallaxes.

Soares \& Bica (2002) used JHK data from the 2MASS Catalogue to study
embedded star clusters associated with NGC~2327 and BRC~27 which is a
bright-rimmed cloud (Sugitani et al. 1991) at $\alpha_{J2000}$ = $7^h
04^m$,  $\delta_{J2000}$ = $ -11^o 23'$. Based
on color-magnitude diagrams they derived a distance of $\sim$ 1.2~kpc
for both clusters, and estimated similar ages of about 1.5 ~Myr. The
mean visual extinction in the direction of the objects were estimated
to be $A_V$~=~5.5 in NGC~2327 and $A_V$~=~6.5 in BRC~27. A subsequent paper
(Soares \& Bica 2003) presented the study of two other
clusters in the nebulae vdB~92 (van den Bergh 1966) and Gy~3-7
(Gyulbudaghian 1984). They adopted the kinematic distance of 1.41 ~kpc
for Gy~3-7 (Wouterloot \& Brand 1989), while the distance estimated for
vdB~92 is 1.5$\pm$0.3~kpc. According to their results, the age derived
for vdB~92 is 5-7~Myr and for Gy~3-7 is $\sim$ 2 ~Myr. The mean
visual extinctions are $A_V$ = 4.4 and $A_V$ = 6.3 mag, respectively.

\section{Distance}

The first determinations of the distance to the CMa OB association
were quite uncertain mainly due to the lack of completeness of the
member content.

Racine (1968) reported photometric and spectroscopic observations of
the CMa~R1 members listed by van den Bergh (1966). The data were used
to determine a mean distance of $690\pm120$ ~pc.  Clari\'a (1974a) used
a color-magnitude diagram based on the early-type stars in the CMa~OB1
region to estimate a distance of 1150$\pm$140 ~pc for the association.
A more detailed study (Clari\'a 1974b) was based on photometric and
spectroscopic data of 44 stars, selected as members or probable
members of CMa~R1.  Clari\'a noted that most of the stars studied by
Racine in CMa~R1 were shown to be CMa~OB1 members and compared the
distance moduli that he obtained with those values estimated by Racine
(1968).  He argues that the discrepancies are due to differences in
the adopted MK types and/or to the disagreement between the absolute
magnitude derived from different hydrogen lines. By correcting these
differences and considering only the true members of CMa R1 (those
which are CMa OB1 members) the derived distance is 1050$\pm$170 ~pc.

Tovmassian et al. (1993) observed 43 early-type stars of the OB
association in CMa and they concluded that these stars compose 3
groups situated at distances 320, 570 and 1100 ~pc. Later, Shevchenko et al. (1999) derived a distance of
1050$\pm$150 ~pc based on photometry  of  88 early-type
stars that were found to be probable members of the CMa R1.

The distribution of young stellar groups in Canis Major was studied
and updated by Kaltcheva \& Hilditch (2000) using uvby$\beta$
photometry of intrinsically luminous OB stars (V$<$9.5mag). They
confirmed a distance of 990$\pm$50 ~pc for CMa~OB1, suggesting an overestimation of the
spectroscopic distances used in the previous studies.

Considering the various estimates, it seems reasonable to recommend
the use of 1000 ~pc for the distance of the CMa clouds.


\begin{figure}[ht!]
\vspace{-1cm}
\plotone{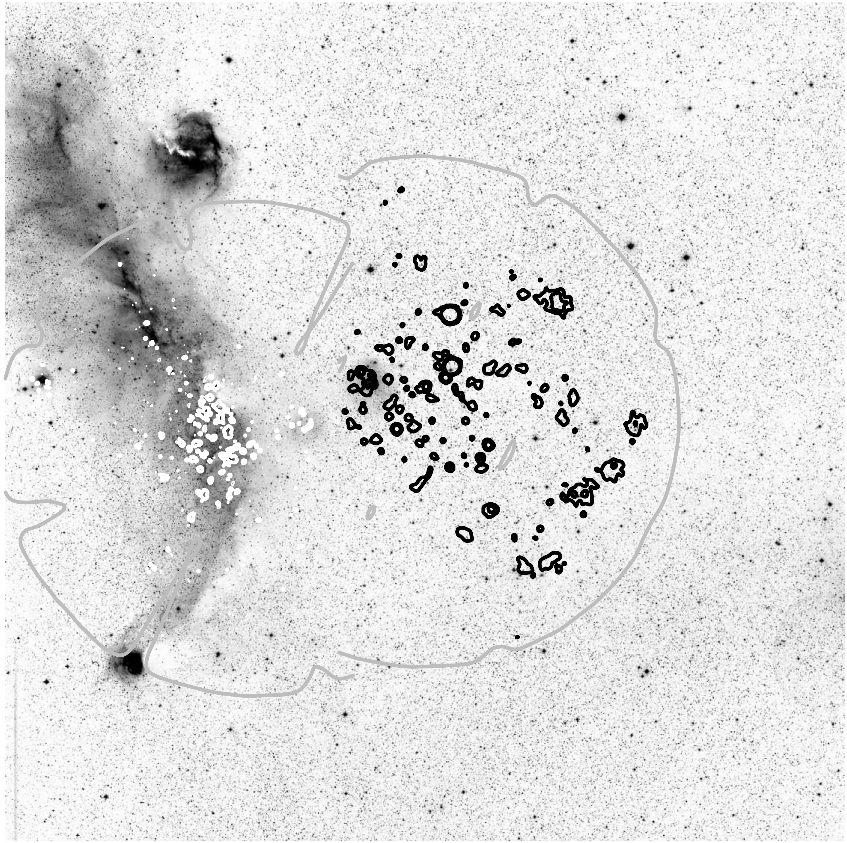}
\vspace{-2.5cm}
\caption{Optical POSS (R) image obtained from a digitized POSS (R) plate made
with the {\it Machine Automatique \`a mesurer pour l'Astronomie}
(MAMA) microdensitometer at Observatoire de Paris superimposed with X-ray contours
obtained from two ROSAT images covering the CMa R1 region. The
sources indicated by white contours (east field)  are
described by Zinnecker \& Preibisch (1994), and the sources in
black contours (west field) by Gregorio-
Hetem, Montmerle, \& Marciotto (2003). The image is 3 $\times$ 3 square degrees. North is up and East is to the left.}
\end{figure}


\begin{figure}[t!]
\plotfiddle{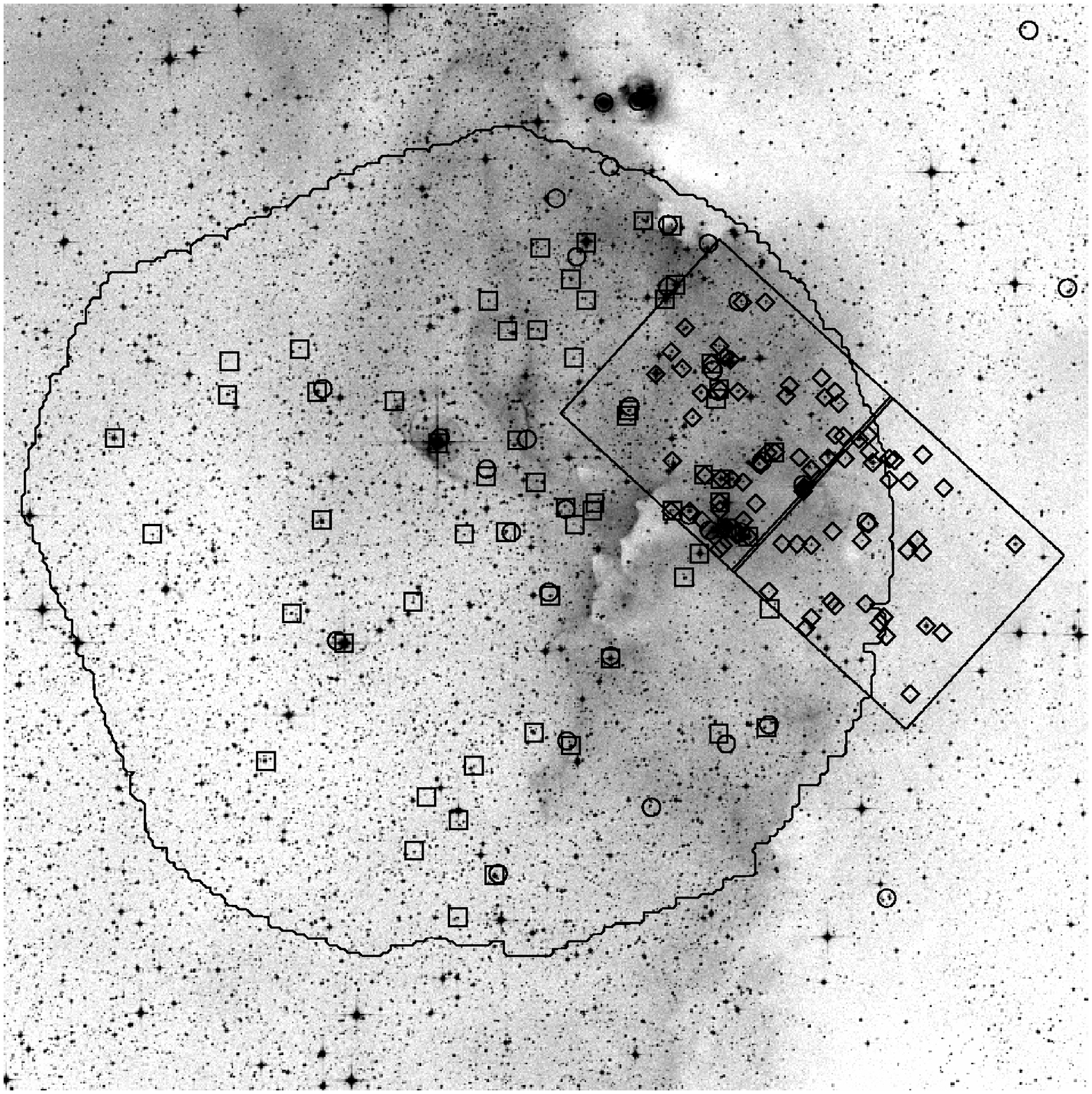}{12cm}{0.0}{50.0}{50.0}{-300.0}{-20.0}
\caption{Recent X-ray observations in CMa R1: the large round area corresponds
to the XMM-Newton field of view, and the two squares at the right indicate the
location of Chandra ACIS-S observations. Open small squares, diamonds and circles
show respectively XMM, Chandra and ROSAT detected sources (extracted from Rojas et al. 2006).
 }
\end{figure}

\section{X-ray Emission}

In order to search for PMS stars, T. Montmerle and collaborators
have studied {\it ROSAT} images of moderately distant molecular
clouds. Most of the selected sources show properties of low- and
intermediate-mass PMS stars: point-like sources are T Tauri and Herbig
Ae/Be stars, while extended emission seems to be due to unresolved
young stellar clusters (Gregorio-Hetem et al. 1998). The 20~ksec {\it
ROSAT} PSPC image obtained for the CMa~R1 region revealed 47 sources
in the central region of the field, with $\sim$80\% having at least one
optical counterpart. Several optical counterparts are emission-line
stars classified as members of CMa~R1 (Shevchenko et al. 1999).  The
bolometric luminosity and spectral type for these optical counterparts
indicate that most of them have $log (L_X/L_{bol})$ in the $-$6 to
$-$4 range.  The 2MASS Catalogue was used to search for sources
coincident with the {\it ROSAT} position error circles, revealing at
least one near-IR counterpart candidate for 90\% of the X-ray
sources. In several cases, multiple candidates were related to a
single extended X-ray emission, possibly corresponding to embedded
clusters.  The evaluation of the color-magnitude (J-H ~vs. ~J)
indicates that 55\% of the sample is consistent with a low-mass young
population, 37\% are intermediate-mass stars and 8\% are massive stars
(Gregorio-Hetem, Montmerle, \& Marciotto 2003).
Additional optical counterparts were identified based on Gemini VRI
data of five X-ray sources in CMa R1. Several counterpart candidates
were found to be associated with each source, showing optical
and near-infrared colors typical of young stars
(Gregorio-Hetem, Rodrigues, \& Montmerle 2007).

An extended feature is also present in the eastern part of
the field, the nature of which is unclear from these  {\it ROSAT}  data: it could be
an unresolved stellar cluster, since the point spread function of the
PSPC is very degraded towards the edge of the field; or a truly
diffuse emission, since it is very precisely located on top of an
optical diffuse feature, which is argued to be either a ``fossil'' HII
region, or an old SNR.

Zinnecker \& Preibisch (1994) performed a PSPC survey of X-ray
emission from Herbig Ae/Be stars in this region, but found no evidence
for extended emission, perhaps due to their short exposures ($\sim$ 5
ksec).  This survey revealed a X-ray luminosity of $(1.4\pm0.7) \times 10^{31}$
erg/s for Z~CMa, but could not detect GU~CMa. Figure ~10 is
a composite of both {\it ROSAT} images obtained of CMa~R1. The field
covering the eastern part is described by Zinnecker \& Preibisch (1994),
and the western part by Gregorio-Hetem, Montmerle, \& Marciotto (2003).
The extended feature mentioned above is not visible in Figure ~10, since it was
overlapped by the image obtained by Zinnecker \& Preibisch (1994).

Improved X-ray observations of the CMa~R1 region were recently
performed with the XMM-Newton telescope towards a 30' field of view
around the extended emission detected by {\it ROSAT}, aiming to elucidate
its origin and to distinguish between unresolved sources and diffuse
emission (Gregorio-Hetem et al. 2008). The new data set reveals the presence of 61 X-ray sources within
the observed field, spatially distributed in clusters previously
unresolved. The XMM data answer the question on the
nature of the extended emission, which appears to be due to unresolved
sources, and seems not to be related to diffuse X-ray emission. An
area of 17 $\times$ 17 arcmin$^2$ in the outer edge of S296 was
covered by Chandra, revealing several additional faint sources in the
field. Using the abovementioned XMM-Newton observations,
 supplemented with Chandra archival data,  Rojas et al. (2006)
identified a total of 135 X-ray sources in the CMa region, see Figure~11.
The correlation of near-IR counterparts with these X-ray
sources indicates that they are probably young stars.

\section{Star Formation induced by a SNR?}

Herbst \& Assousa (1977) reported evidence for star formation in
CMa~R1 induced by a supernova explosion. The suggestion is based on:
{\it(i)} the form of a large scale ring of emission nebulosity with a
diameter of about $3^o$, defined by S296 and fainter nebulae to the
north and east, which could be an old supernova remnant; {\it(ii)} the
absence of luminous stars at the ring center, which is  confirmed by the
far ultraviolet image of the field by Henry (1988);
{\it(iii)} an expanding neutral hydrogen shell coinciding with the optical
feature and evident in the H I maps of Weaver \& Williams (1974); {\it(iv)}
the discovery in CMa~OB1 of a runway star, HD~54662, that could be
associated with the event that produced the SNR.  Velocity information
for the components of the CMa~OB1/R1 complex was compiled by Herbst \&
Assousa (1977) from different catalogues of H I, Ca II, CO and H II
velocities to estimate the mean radial velocity of the
association. They verified that the early-type star HD~54662 lies
within the ring, but it is offcenter and has a radial velocity
differing by about 30 ~km/s from the velocity inferred for the
association. Based on its peculiar velocity, HD~54662 was considered a
possible runaway star, but confirmation depends on an accurate
determination of the proper motion.

Herbst \& Assousa (1977) compared the SNR properties with models of
SNRs evolving in a uniform medium to derive an age of 5$\times 10^5$
yrs for the supernova shell. A study of the optical and infrared
properties of the stars in CMa~R1 by Herbst, Racine, \& Warner (1978)
provided an estimate of a similar age for most of the members of the R
association, compatible with the suggestion of star formation
triggered by a supernova in this region. On the other hand, Shevchenko et al.
(1999) concluded that only a minor fraction of the young stars in CMa R1
could have been created by the mechanism suggested by Herbst \& Assousa,
 since the majority of them have ages that are
much older than that required by this scenario.

Radio-continuum observations obtained by Nakano et al. (1984) indicate
that the OB stars are sufficient to ionize the emission nebulosity.
The number of Lyman continuum photons emitted by seven of the brightest
members (HD~54662,
53975,
53974,
53367,
54879,
53456,
53755)
 was computed to determine the total photon flux from OB stars.
The comparison with the thermal spectrum exhibited by S296 confirms
the ionization balance between OB stars and the ionized gas in this
region. A SNR
could have produced the ring morphology and triggered the episode of
star formation, but the remnant would be so old that no trace of
non-thermal emission would be expected today.

Vrba et al. (1987) performed a linear polarization survey in the
direction of CMa~R1 in order to test the hypothesis of star formation
driven by a shock wave. The polarizations were found to be
inconsistent with a scenario of quiescent cloud collapse. These
results are compatible with a simple model of the compression by a
supernova-induced spherical shock front of an initially uniform
interstellar magnetic field, or alternatively, the shell-like
structure could be produced by stellar winds.

The gas velocities in the region were also estimated by Reynolds \&
Ogden (1978) measuring the H$\alpha$, [NII] and [OIII] lines, which
were used to confirm that the large shell is in expansion with a
velocity of about 13 ~km/s. The authors also noted that the ultraviolet
fluxes of two hot stars, HD~54662 and HD~53975, located within the ring
are sufficient to account for the temperature and ionization of the
shell. Based on these results, Reynolds \& Ogden suggested strong
stellar winds, or an evolving HII region, as alternatives to the SNR
hypothesis, as also suggested by Blitz (1980) and Pyatunina \& Taraskin
(1986).

More recently, Comer\'on et al. (1998) used proper motions measured by
{\it Hipparcos } to reveal a clearly organized expanding pattern in
CMa~OB1 showing LSR velocity of $\sim$15~km/s. Their results on the
magnitude and spatial arrangement of the expanding motions suggest
that very energetic phenomena are responsible for the formation of the
OB association. A rough determination of the center at ({\em l,b}) = (226.5,
--1.6) and age $\sim$ 1.5 ~Myr was carried out for the expanding
structure. A new runaway star HD~57682 was discovered, showing a motion
directly away from the derived center of expansion, which supports the
scenario suggested by Herbst \& Assousa.


\vspace{0.5cm}

{\bf Acknowledgements.}  This chapter is based on the original work
``The Canis Majoris OB1 Associations" published by George Herbig 15
years ago in the ESO book ``Low Mass Star Formation in Southern
Molecular Clouds". I would like to express my gratitude to Dr. Herbig
for his kindness to provide some of the original figures and for his
valuable comments on the manuscript. A special thank goes to Bo
Reipurth for his suggestions and for critiquing drafts of
the manuscript. I also thank Mario van den Ancker, the referee, for
constructive comments that improved the presentation of the chapter;
John P. Gleason for kindly providing Figure~1; and Steve Rodney for
useful suggestions on the layout of tables.

This work has made use of the SIMBAD, VizieR, and
Aladin databases operated at CDS, Strasbourg, France. Partial financial support from FAPESP
(Procs. No. 2001/09018-2 and No. 2005/00397-1) is also acknowledged.


\end{document}